# Data-driven Control of an LCC HVDC System for Real-time Frequency Regulation

Do-Hoon Kwon, *Member*, *IEEE* and Young-Jin Kim, *Member*, *IEEE*

*Abstract*—Recent advances in data sensing and processing technologies enable data-driven control of high-voltage direct-current (HVDC) systems for improving the operational stability of interfacing power grids. This paper proposes an optimal data-driven control strategy for an HVDC system with line-commutated converters (LCCs), wherein the dc-link voltage and current are optimally regulated at distinct HVDC terminals to improve frequency regulation (FR) in both rectifier- and inverter-side grids. Each HVDC converter is integrated with feedback loops for regulation of grid frequency and dc-link voltage in a localized manner. For optimal FR in both-side grids, a data-driven model of the HVDC-linked grids is then developed to design a data-driven linear quadratic Gaussian (LQG) regulator, which is incorporated with the converter feedback loops. Case studies on two different LCC HVDC systems are performed using the data-driven models, which are validated via comparisons with physics-based models and comprehensive MATLAB/SIMULINK models. The results of the case studies confirm that the optimal data-driven control strategy successfully exploits the fast dynamics of HVDC converters; moreover, cooperation of the HVDC system and synchronous generators in both-side grids is achieved, improving real-time FR under various HVDC system specifications, LQG parameters, and inertia emulation and droop control conditions.

*Index Terms*— data-driven control, dc-link voltage and current, high-voltage direct-current systems, line-commutated converters, linear quadratic Gaussian, real-time frequency regulation.

## I. INTRODUCTION

HIGH-voltage direct-current (HVDC) systems have been widely used in modern power grids, and can provide long transmission distances, asynchronous grid connection, and fast and flexible converter control [1]. In many projects worldwide, wind farms (WFs) ranging in size from 500 MW to 1,500 MW have been connected to distant load centers via HVDC systems, using the mature and trusted line-commutated converter (LCC) technology [2], [3]. For example, an LCC HVDC system has been installed in the Midwest area of the United States to transfer approximately 25% of the power generated by wind turbines, with a total capacity of 16 GW, to the Eastern Interconnection [4].

Many studies (e.g., [5]–[15]) on dynamic control of LCC HVDC systems have investigated methods for alleviating the impact of intermittent wind power on grid operations, facilitating grid interconnections via HVDC systems and supporting real-time frequency regulation (FR). For example, in [5], the rectifier firing angle or, equivalently, dc current was controlled to improve FR in the rectifier-side grid connected to a WF. In [6], the mechanical kinetic energy of a WF in the rectifier-side grid was exploited to enhance FR in the inverter-side grid. In [7], the effect of a dc inertia constant on FR was analyzed when a weak grid was linked to both a WF and an LCC HVDC system. A fuzzy-based strategy was adopted in [8] to improve frequency damping and voltage oscillations in the inverter-side grid. However, in [5]–[8], HVDC system control aimed to support FR only in either the inverter- or rectifier-side grid, because the grid on the other side was assumed to include only WFs, rather than synchronous generators and loads. In recent studies (e.g., [9]–[11]), FR has been achieved at both terminals of an LCC HVDC system. For example, in [9], firing angle control and WF droop control were discussed using a four-state nonlinear model reflecting the coupling between the rectifier- and inverter-side grids. Given the grid coupling, optimal control of HVDC systems was discussed in [10] and [11], where linear quadratic (LQ) regulators were adopted to enhance voltage angle and rotor angle oscillation damping, respectively.

In the previous studies, the dc-link voltage references were commonly held constant at the rated values. In other words, the LCC HVDC systems supported FR in single- or both- side grids by controlling the dc currents only. Under normal conditions, LCCs can operate with under- and over-voltages of the dc link for a short period of time, as reported in field tests [12], [13]. Recently, a few studies have been conducted on time-varying control of the dc-link voltages of LCC HVDC systems. For example, in [14] and [15], the capabilities of LCC HVDC systems for short-term power transfer and reactive power control, respectively, were improved by regulating the dc voltages dynamically at the rectifier and inverter terminals.

In [5]–[15], control strategies were implemented using physics-based models of HVDC systems and generators. In general, physics-based modeling requires a number of device parameters, most of which remain unidentified or vary widely by device type, size, and structure. For practical applications, control strategies should be accompanied by techniques for parameter estimation and optimal gain tuning, which are usually time-consuming and difficult to apply to large-scale power networks that are already in daily service. Due to recent advances in sensing and data technologies, increasing amounts of grid operating data, for example, on frequency, net load, and WF power generation [16], [17] can be collected and processed in real time, resolving the challenges associated with physics-model-based control. For example, in [18] and [19], model-free adaptive control (MFAC) was applied to ac-dc inter-linking converters and inverter-based generators, respectively. In [20] and [21], reinforcement learning (RL) was used for FR in isolated- and interconnected-grids, respectively. The MFAC requires the input-and-output function of power equipment to satisfy a generalized Lipschitz condition. In RL, the optimal policy can be obtained only after repeated trials while searching over large state and action spaces, which makes the application of RL to real power systems difficult. In

Manuscript received Nov. 17, 2019 (corresponding author: Y. Kim).
D. H. Kwon is with Korea Electrotechnology Research Institute (KERI), Uiwang-si, Gyeonggi-do, 16029, Korea (e-mail: dhkwon@keri.re.kr).
Y. Kim is with the Department of Electrical Engineering, Pohang University of Science and Technology, Pohang, Gyungbuk 37673, Korea (e-mail: powersys@postech.ac.kr).

[22] and [23], the observer/Kalman filter identification (OKID) and eigensystem realization algorithm (ERA) were discussed for data-driven control of general linear and nonlinear systems.

This paper proposes an optimal data-driven control strategy for an LCC HVDC system, wherein the dc-link voltage and current are optimally regulated at distinct HVDC terminals via primary (PFC) and secondary frequency control (SFC) to improve real-time FR in both rectifier- and inverter-side grids. To attain PFC, each HVDC converter is integrated with feedback loops for $P_{dc}$-$f$ and $P_{dc}$-$V_{dc}$ droop control, as well as inertial response emulation (IRE). A linear quadratic Gaussian (LQG) regulator is then designed to achieve optimal SFC of the HVDC system and synchronous generators, given the IRE and PFC schemes, thereby minimizing the weighted sum of deviations of the dc-link voltage and grid frequencies in the rectifier- and inverter-side grids. To improve the practical applicability of the optimal SFC scheme, a data-driven model of the HVDC-linked grids is developed using the OKID and ERA, considering modeling complexity and the computational effort required to process data measurements. This enables the LQG regulator to be purely data-driven. Simulation case studies are carried out for two LCC HVDC systems (i.e., a real HVDC system and a CIGRE benchmark system). The data-driven models are validated via comparisons with the corresponding physics-based models and comprehensive MATLAB/SIMULINK models. The results of the case studies confirm the effectiveness of the proposed data-driven control strategy for improving real-time FR in both-side grids under various conditions, characterized by the HVDC system specifications; the LQG weighting factors; the inertia emulation and droop control; and the time-varying load demand and wind power generation.

The main contributions of this paper are summarized below:
• To the best of our knowledge, this is the first study to develop a data-driven strategy for optimal real-time control of an LCC HVDC system in coordination with synchronous generators, based on a data-driven dynamic model of HVDC-linked grids and a data-driven LQG regulator.
• The data-driven model of the HVDC-linked grids is developed using input and output data measured during normal operation. Singular value decomposition (SVD) and Kalman filtering are also applied to facilitate real-time model identification and hence the data-driven control strategy.
• The IRE and PFC schemes are applied to exploit the fast response of HVDC converters. The data-driven LQG regulator is then designed to optimally adjust the dc-link voltage and current at distinct HVDC terminals, supporting improvements in the load sharing of generators and hence FR in both the rectifier- and inverter-side grids.

## II. PROPOSED FREQUENCY CONTROL IN HVDC-LINKED GRIDS

Fig. 1 shows a simplified schematic diagram of the proposed real-time FR in the inverter- and rectifier-side grids linked via an LCC HVDC system. The grid frequency deviations $\Delta f_r$ and $\Delta f_i$ on both sides of the HVDC system result from wind power fluctuation $\Delta P_w$ and load demand variations $\Delta P_{lr}$ and $\Delta P_{li}$ on either side. In the proposed FR strategy, the HVDC system controls the dc voltage and current (i.e., $\Delta V_{dcr}$ and $\Delta I_{dci}$) at the rectifier and inverter terminals, respectively, in response to $\Delta f_r$ and $\Delta f_i$. This leads to variations in the dc-link voltage $\Delta V_{dc}$ and the power transferred via the dc link: i.e., $\Delta P_{dcr}$ flowing into the rectifier and $\Delta P_{dci}$ flowing from the inverter. Dynamic control of the HVDC system is achieved in coordination with the frequency response of synchronous generators in both-side grids, where the total power outputs $\Delta P_{gr}$ and $\Delta P_{gi}$ are also adjusted in real time according to $\Delta f_r$ and $\Delta f_i$, respectively.

The HVDC system and generators are equipped with droop controllers to achieve active power sharing and hence stabilize $f_r$ and $f_i$ in a localized manner. The HVDC system has additional droop controllers to assign the references of $\Delta V_{dcr}$ and $\Delta I_{dci}$ separately in response to $\Delta V_{dc}$, so that $V_{dc}$ can be stabilized according to the individual operating conditions of the rectifier and inverter. This enables $V_{dc}$ to be controlled within an acceptable range, ensuring robust support of the HVDC system for active power sharing of the generators. As shown in Fig. 1, each HVDC converter is also integrated with a local controller to support the inertial response of the interfacing grid, so that the fast dynamics of the converters can be further exploited for real-time FR. Moreover, an LQG regulator is designed to achieve the optimal, coordinated SFC of the HVDC system and generators, where the references of $\Delta V_{dcr}$, $\Delta I_{dci}$, $\Delta P_{gr}$, and $\Delta P_{gi}$ are optimally determined in real time to restore $f_i$, $f_r$, and $V_{dc}$ back to their nominal values.

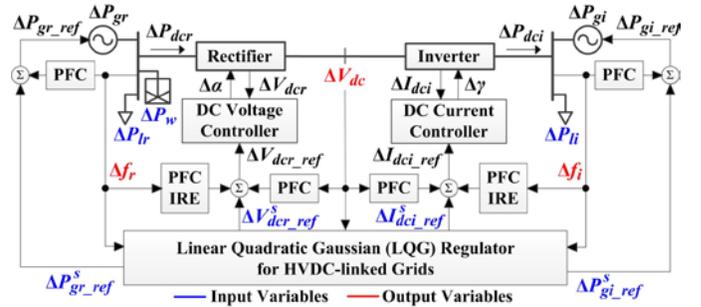

Fig. 1. Schematic diagram of the proposed optimal control strategy for an LCC HVDC system to support real-time FR in rectifier- and inverter-side grids.

### A. DC Voltage and Current Control of LCC HVDC System

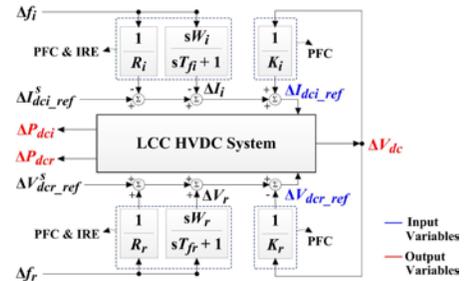

Fig. 2. An LCC HVDC system with the IRE, PFC, and SFC schemes for controlling the dc-link voltage and current in response to $\Delta f_i$, $\Delta f_r$, and $\Delta V_{dc}$.

Fig. 2 shows a schematic diagram of the proposed control strategy for an LCC HVDC system, as briefly discussed above. The input variables are defined as $\Delta V_{dcr\_ref}$ and $\Delta I_{dci\_ref}$, so that the responsibilities to control $\Delta V_{dcr}$ and $\Delta I_{dci}$ can be kept distinct and assigned to separate converters to support FR in both-side grids. This also allows the output variables $\Delta P_{dcr(i)}$ and $\Delta V_{dc}$ to be readily estimated, thus facilitating practical implementation of the proposed strategy.

For the PFC, the droop controllers of the inverter and rectifier are designed with proportional gains of $1/R_{i(r)}$ to stabilize $\Delta f_{i(r)}$, respectively, in a localized manner. For the stabilization of $\Delta f_{i(r)}$, the IRE scheme is also implemented using a derivative feedback controller with a gain of $W_{i(r)}$ and a low-pass filter with a time constant of $T_{fi(r)}$ [6]. Similarly, droop controllers with gains of $1/K_{i(r)}$ are established to stabilize $\Delta V_{dc}$. The gains for the IRE and PFC schemes can be readily determined without comprehensive analysis of the interactions between the HVDC system and generators. The gain tuning of the control schemes is also marginally affected by the specific models of the dc link, converters, and inner feedback loops for firing angle control.

### B. Output Power Control of Gas-turbine Generator

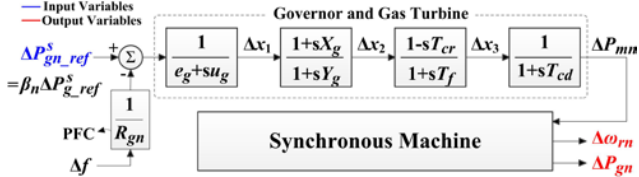

Fig. 3. A gas-turbine synchronous generator with the PFC and SFC schemes.

In this study, a gas-turbine-driven synchronous generator [24] is considered to be operating as a grid-forming unit in the HVDC-linked grids. The synchronous generator is capable of directly regulating the rotor's speed $\Delta \omega_r$ and hence the grid frequency, given the instantaneous imbalance between the mechanical power input $\Delta P_m$ and electrical power output $\Delta P_g$ for the real-time power $\Delta P_{dci(r)}$ transferred via the HVDC system. As shown in Fig. 3, the synchronous generator is equipped with feedback loops for the PFC and SFC. The PFC is achieved using a proportional gain of $1/R_{gn}$. As in case of the HVDC system, the PFC can be readily implemented in a localized manner, without detailed modeling or analysis of the operations of the HVDC system and generators. For the SFC, a participation factor $\beta_n$ is considered to calculate the required variation in power output of a generating unit $n$, given the variation in total power output reference $\Delta P^S_{g\_ref}$ for all generators on each side of the HVDC system.

### III. OPTIMAL DATA-DRIVEN FREQUENCY REGULATION

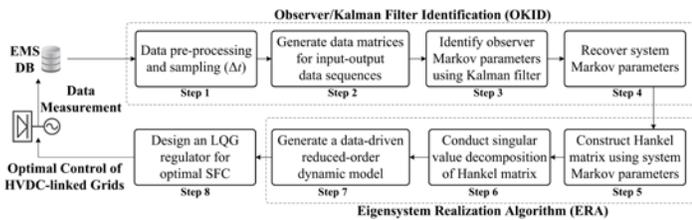

Fig. 4. Data-driven modeling and control of the HVDC-linked grids.

For the optimal SFC of the HVDC-linked grids, it is necessary to develop small-signal models of the LCC HVDC system and synchronous generators, discussed in Sections II-A and II-B, respectively. A data-driven, reduced-order modeling (DRM) approach is adopted here to develop a dynamic model of the HVDC-linked grids, removing the need to estimate the large number of modeling parameters required for a physics-based modeling approach, as discussed in Appendix. This facilitates practical application of the proposed optimal data-driven control strategy of the HVDC system. The DRM approach consists of the OKID and ERA, as shown in Fig. 4.

### A. System Identification Using Observer/Kalman Filter

Briefly, the OKID method aims to estimate the impulse response of a dynamic system based on its input and output data, as measured under normal operating conditions. A state-space model of the HVDC-linked grids with the IRE and PFC schemes, discussed in Section II, can be established as:

$$d\mathbf{X}(t)/dt = \mathbf{A} \cdot \mathbf{X}(t) + \mathbf{B} \cdot [\mathbf{r}(t) \quad \mathbf{w}(t)]^T, \quad (1)$$

$$\mathbf{Y}(t) = \mathbf{C} \cdot \mathbf{X}(t) + \mathbf{D} \cdot [\mathbf{r}(t) \quad \mathbf{w}(t)]^T, \quad (2)$$

where 
$$\mathbf{r}(t) = [\Delta P^S_{gi\_ref}, \Delta P^S_{gr\_ref}, \Delta I^S_{dci\_ref}, \Delta V^S_{dcr\_ref}]^T, \quad (3)$$

$$\mathbf{w}(t) = [\Delta P_{li}, \Delta P_{lr} - \Delta P_w]^T, \text{ and} \quad (4)$$

$$\mathbf{Y}(t) = \left[ \Delta f_i, \Delta f_r, \Delta V_{dc}, \int \Delta f_i \, dt, \int \Delta f_r \, dt, \int \Delta V_{dc} \, dt \right]^T. \quad (5)$$

Note that the OKID method is purely data-driven; therefore, coefficient matrices of $\mathbf{A}$, $\mathbf{B}$, $\mathbf{C}$, and $\mathbf{D}$ need not be identified. For a sampling time of $\Delta t = T_s$, the discrete-time model of the HVDC-lined grids, corresponding to (1)–(5), is represented as:

$$\mathbf{X}_{k+1} = \mathbf{A}_d \cdot \mathbf{X}_k + \mathbf{B}_d \cdot [\mathbf{r}_k \quad \mathbf{w}_k]^T, \quad (6)$$

$$\mathbf{Y}_k = \mathbf{C}_d \cdot \mathbf{X}_k + \mathbf{D}_d \cdot [\mathbf{r}_k \quad \mathbf{w}_k]^T, \quad (7)$$

where $\mathbf{A}_d = e^{\mathbf{A}T_s}$, $\mathbf{B}_d = \int_0^{T_s} e^{\mathbf{A}\tau} \mathbf{B} \, d\tau$, $\mathbf{C}_d = \mathbf{C}$, and $\mathbf{D}_d = \mathbf{D}$. (8)

For the discrete-time system (6)–(8), the impulse response is then obtained in the form of *system Markov parameters* as:

$$\mathbf{Y}^\delta_0 = \mathbf{D}_d, \ \mathbf{Y}^\delta_1 = \mathbf{C}_d \mathbf{B}_d, \ \cdots, \text{ and } \mathbf{Y}^\delta_k = \mathbf{C}_d \mathbf{A}^{k-1}_d \mathbf{B}_d, \quad (9)$$

when the Kronecker unit pulse sequence $\boldsymbol{\delta}_k$ in (10) is assigned to the input $\mathbf{U}_k = [\mathbf{r}_k \quad \mathbf{w}_k]^T$.

$$\boldsymbol{\delta}_k = \begin{cases} \mathbf{I} & \text{for } k = 0, \\ \mathbf{O} & \text{for } k \geq 1. \end{cases} \quad (10)$$

The system Markov parameters (or, equivalently, the impulse response sequence) in (9) contain information on the dynamics of the discrete-time system model (6)–(8) and, hence, the continuous-time model (1)–(5), which is required for optimal SFC of the HVDC-linked grids, as discussed in Section III-C. It is difficult to apply $\boldsymbol{\delta}_k$ to real HVDC systems and synchronous generators. Therefore, the system Markov parameters need to be estimated by measuring the output data $\mathbf{Y}_k$ in response to a general input signal $\mathbf{U}_k$. In normal operation with a zero initial condition $\mathbf{X}_{k=0} = \mathbf{O}$, $\mathbf{Y}_k$ for any $\mathbf{U}_k$ is given by:

$$\begin{aligned}
\mathbf{Y}_0 &= \mathbf{D}_d \mathbf{U}_0, \\
\mathbf{Y}_1 &= \mathbf{C}_d \mathbf{B}_d \mathbf{U}_0 + \mathbf{D}_d \mathbf{U}_1 = \mathbf{Y}^\delta_1 \mathbf{U}_0 + \mathbf{Y}^\delta_0 \mathbf{U}_1, \\
&\vdots \\
\mathbf{Y}_k &= \mathbf{Y}^\delta_k \mathbf{U}_0 + \mathbf{Y}^\delta_{k-1} \mathbf{U}_1 + \cdots + \mathbf{Y}^\delta_0 \mathbf{U}_k.
\end{aligned} \quad (11)$$

In compact form, (11) is represented as:

$$\underbrace{[\mathbf{Y}_0 \ \mathbf{Y}_1 \ \cdots \ \mathbf{Y}_m]}_{Y} = \underbrace{[\mathbf{Y}^\delta_0 \ \mathbf{Y}^\delta_1 \ \cdots \ \mathbf{Y}^\delta_m]}_{Y^\delta} \cdot \underbrace{\begin{bmatrix} \mathbf{U}_0 & \mathbf{U}_1 & \cdots & \mathbf{U}_m \\ \mathbf{O} & \mathbf{U}_0 & \cdots & \mathbf{U}_{m-1} \\ \vdots & \vdots & \ddots & \vdots \\ \mathbf{O} & \mathbf{O} & \cdots & \mathbf{U}_0 \end{bmatrix}}_{U}, \quad (12)$$

where $m$ is the number of sampled data. As shown in (12), the system Markov parameters $Y^\delta$ can be estimated from the general data sequences of $\mathbf{U}_k$ and $\mathbf{Y}_k$.

It is possible that $U$ is un-invertible or its inverse matrix is ill-conditioned, for example, due to sensing noises. The inversion can be computationally expensive due to the large number of data sequences. To address these issues, a state observer is established using a Kalman filter [25] with a gain of $\mathbf{K}_f$ as:

$$\hat{\mathbf{X}}_{k+1} = (\mathbf{A}_d - \mathbf{K}_f \mathbf{C}_d) \cdot \hat{\mathbf{X}}_k + [\mathbf{B}_d - \mathbf{K}_f \mathbf{D}_d, \ \mathbf{K}_f] \cdot [\mathbf{U}_k \ \mathbf{Y}_k]^T. \quad (13)$$

For (13), the observer Markov parameters $\bar{Y}^\delta$ can be estimated using the general data sequences of $\mathbf{U}_k$ and $\mathbf{Y}_k$ as:

$$\underbrace{[\mathbf{Y}_0 \ \mathbf{Y}_1 \ \cdots \ \mathbf{Y}_m]}_{Y} = \underbrace{[\bar{\mathbf{Y}}_0^\delta \ \bar{\mathbf{Y}}_1^\delta \ \cdots \ \bar{\mathbf{Y}}_m^\delta]}_{\bar{Y}^\delta} \cdot \underbrace{\begin{bmatrix} \mathbf{U}_0 & \mathbf{U}_1 & \cdots & \mathbf{U}_l & \cdots & \mathbf{U}_m \\ \mathbf{O} & \mathbf{V}_0 & \cdots & \mathbf{V}_{l-1} & \cdots & \mathbf{V}_{m-1} \\ \vdots & \vdots & \ddots & \vdots & \ddots & \vdots \\ \mathbf{O} & \mathbf{O} & \cdots & \mathbf{V}_0 & \cdots & \mathbf{V}_{m-l} \end{bmatrix}}_{V}, \quad (14)$$

where $\mathbf{V}_k = [\mathbf{U}_k^T \ \mathbf{Y}_k^T]^T$, $\bar{\mathbf{Y}}_0^\delta = \mathbf{D}$, and $\bar{\mathbf{Y}}_k^\delta = [\bar{\mathbf{Y}}_k^{\delta(1)} \ \bar{\mathbf{Y}}_k^{\delta(2)}]$ for $k \geq 1$. Although $V$ in (14) resembles $U$ in (12), the size of $V$ can be made significantly smaller than that of $U$ by choosing the appropriate number $l$ of the observer Markov parameters to estimate. This facilitates the inversion of $V$. Note that $l$ can still be sufficiently large to ensure the existence of $V^{-1}$.

Since $\bar{Y}^\delta = Y \cdot V^T \cdot (V \cdot V^T)^{-1}$ in (14), $\mathbf{Y}_k^\delta$ in (12) can be reconstructed using (15) and (16), rather than using $Y^\delta = Y \cdot U^{-1}$.

$$\mathbf{Y}_0^\delta = \bar{\mathbf{Y}}_0^\delta = \mathbf{O}, \quad (15)$$

$$\mathbf{Y}_k^\delta = \bar{\mathbf{Y}}_k^{\delta(1)} + \sum_{i=1}^{k} \bar{\mathbf{Y}}_i^{\delta(2)} \mathbf{Y}_{k-i}^\delta \quad \text{for } k \geq 1. \quad (16)$$

### B. Data-driven, Reduced-order Dynamic Modeling

The ERA takes $\mathbf{Y}_k^\delta$ in (15) and (16) as inputs to develop a dynamic model whose impulse response is the same as that of (1)–(5). Specifically, the Hankel matrix $\mathbf{H}$ is formed by stacking time-delayed values of $\mathbf{Y}_k^\delta$ for $1 \leq k \leq 2p-1 \leq m$ as:

$$\mathbf{H} = \begin{bmatrix} \mathbf{Y}_1^\delta & \mathbf{Y}_2^\delta & \cdots & \mathbf{Y}_p^\delta \\ \mathbf{Y}_2^\delta & \mathbf{Y}_3^\delta & \cdots & \mathbf{Y}_{p+1}^\delta \\ \vdots & \vdots & \ddots & \vdots \\ \mathbf{Y}_p^\delta & \mathbf{Y}_{p+1}^\delta & \cdots & \mathbf{Y}_{2p-1}^\delta \end{bmatrix}. \quad (17)$$

The SVD algorithm [26] is then applied to $\mathbf{H}$, as shown in (18) and (19), so that only the dominant temporal pattern in $\mathbf{Y}_k^\delta$ is considered to construct a low-order dynamic model of the HVDC-linked grids. This mitigates the effects of sensing noises and reduces the model construction time, thereby facilitating practical implementation of the optimal SFC scheme, as discussed in Section III-C.

$$\mathbf{H} = \mathbf{L}\mathbf{S}\mathbf{R}^* = [\tilde{\mathbf{L}} \ \mathbf{L}_t]\begin{bmatrix} \tilde{\mathbf{S}} & \mathbf{O} \\ \mathbf{O} & \mathbf{S}_t \end{bmatrix}\begin{bmatrix} \tilde{\mathbf{R}}^* \\ \mathbf{R}_t^* \end{bmatrix}, \quad (18)$$

$$\approx \tilde{\mathbf{H}} = \tilde{\mathbf{L}}\tilde{\mathbf{S}}\tilde{\mathbf{R}}^* = \begin{bmatrix} \mathbf{Y}_1^\delta & \mathbf{Y}_2^\delta & \cdots & \mathbf{Y}_r^\delta \\ \mathbf{Y}_2^\delta & \mathbf{Y}_3^\delta & \cdots & \mathbf{Y}_{r+1}^\delta \\ \vdots & \vdots & \ddots & \vdots \\ \mathbf{Y}_r^\delta & \mathbf{Y}_{r+1}^\delta & \cdots & \mathbf{Y}_{2r-1}^\delta \end{bmatrix}. \quad (19)$$

In (18), $\mathbf{L}$ and $\mathbf{R}$ represent the left- and right-singular vectors of $\mathbf{H}$, respectively, and the diagonal elements of $\mathbf{S}$ are the real, non-negative Hankel singular values (HSVs), which are ordered from largest to smallest. In other words, $\tilde{\mathbf{S}}$ contains the $r$ ($< p$) large HSVs to be retained, whereas $\mathbf{S}_t$ includes the ($p$–$r$) small HSVs to be truncated. The truncated Hankel matrix $\tilde{\mathbf{H}} = \tilde{\mathbf{L}}\tilde{\mathbf{S}}\tilde{\mathbf{R}}^*$ in (19) can still capture the dominant dynamics in $\mathbf{H}$.

Fig. 5(a) and (b) show the magnitudes of the HSVs and the corresponding normalized cumulative energy [26], respectively, for $1 \leq r \leq p$, when $\mathbf{H}$ is established with an arbitrarily large value of $p$ for the test condition of the HVDC-linked grids, discussed in Section IV-A (see Table I). For $r = 16$, $\tilde{\mathbf{H}}$ in (19) captures over 99.9% of the total energy contained in $\mathbf{H}$. This indicates that a $16_{th}$-order data-driven model can successfully reflect the operating characteristics of the HVDC- linked grids. Note that the physics-based modeling approach, discussed in Appendix, results in a $204_{th}$-order small-signal model.

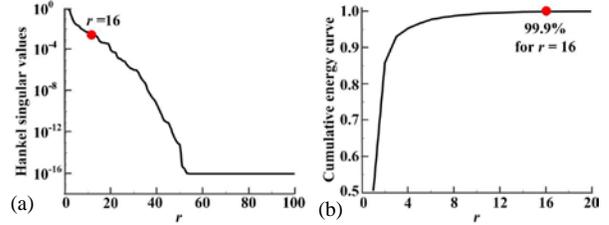

Fig. 5. (a) Hankel singular values and (b) cumulative energy curve of $\mathbf{H}$ for the HVDC-linked grids. The first 16 HSVs contain over 99.9% of the energy.

Using $\tilde{\mathbf{H}} = \tilde{\mathbf{L}}\tilde{\mathbf{S}}\tilde{\mathbf{R}}^*$ in (19), the $r_{th}$-order, data-driven dynamic model of the HVDC-linked grids can be developed [26] as:

$$\hat{\mathbf{X}}_{k+1} = \tilde{\mathbf{A}}_d \hat{\mathbf{X}}_k + \tilde{\mathbf{B}}_d \mathbf{U}_k, \quad (20)$$

$$\mathbf{Y}_k = \tilde{\mathbf{C}}_d \hat{\mathbf{X}}_k + \tilde{\mathbf{D}}_d \mathbf{U}_k, \quad (21)$$

where $\tilde{\mathbf{A}}_d = \tilde{\mathbf{S}}^{-1/2}\tilde{\mathbf{L}}^* \tilde{\mathbf{H}}'\tilde{\mathbf{R}}\tilde{\mathbf{S}}^{-1/2}$, $\tilde{\mathbf{B}}_d = \tilde{\mathbf{S}}^{-1/2}\tilde{\mathbf{R}}^*\begin{bmatrix} \mathbf{I}_z & \mathbf{O} \\ \mathbf{O} & \mathbf{O} \end{bmatrix}$, (22)

$$\tilde{\mathbf{C}}_d = \begin{bmatrix} \mathbf{I}_v & \mathbf{O} \\ \mathbf{O} & \mathbf{O} \end{bmatrix}\tilde{\mathbf{L}}\tilde{\mathbf{S}}^{-1/2}, \quad \tilde{\mathbf{D}}_d = \mathbf{Y}_0^\delta. \quad (23)$$

In (22) and (23), $v$ and $z$ are the numbers of input and output variables, respectively: i.e., $v = 6$ and $z = 6$ for (1)–(5). Moreover, $\tilde{\mathbf{H}}'$ in (22) is a unit-time-shifted matrix of $\tilde{\mathbf{H}}$. To complete the data-driven modeling, the discrete-time model (20)–(23) is converted into a continuous-time model [27] as:

$$d\tilde{\mathbf{X}}(t)/dt = \tilde{\mathbf{A}} \cdot \tilde{\mathbf{X}}(t) + \tilde{\mathbf{B}} \cdot [\mathbf{r}(t) \ \mathbf{w}(t)]^T + \mathbf{G}_d, \quad (24)$$

$$= \tilde{\mathbf{A}} \cdot \tilde{\mathbf{X}}(t) + \tilde{\mathbf{B}}_r \cdot \mathbf{r}(t) + \tilde{\mathbf{B}}_w \cdot \mathbf{w}(t) + \mathbf{G}_d,$$

$$\mathbf{Y}(t) = \tilde{\mathbf{C}} \cdot \tilde{\mathbf{X}}(t) + \tilde{\mathbf{D}} \cdot [\mathbf{r}(t) \ \mathbf{w}(t)]^T + \mathbf{G}_n, \quad (25)$$

which is generalized by including stochastic disturbances $\mathbf{G_d}$ and sensor noise $\mathbf{G_n}$. Note that the coefficient matrices $\tilde{\mathbf{A}}$, $\tilde{\mathbf{B}}_r$, $\tilde{\mathbf{B}}_w$, $\tilde{\mathbf{C}}$, and $\tilde{\mathbf{D}}$ are calculated using only the measurements of the general input and output data of the HVDC-linked grids under normal operating conditions. This implies that the optimal SFC, as discussed in Section III-C, can be successfully achieved only using the sensed data on the operations of the HVDC system and synchronous generators, thereby significantly enhancing the practical applicability of the proposed FR strategy. It also should be noted that the IRE and PFC schemes for the HVDC system and generators, discussed in Section II, are reflected in the data-driven state-space model (24), (25). Furthermore, the input and output variables of (24) and (25) remain the same as those in (1)–(5). Therefore, the data-driven model (24), (25) also can be applied to conventional output feedback controllers such as a PI controller, as discussed in Section IV-A.

### C. Optimal SFC of the LCC-HVDC-linked Grids

Fig. 6 shows an LQG regulator for the optimal SFC of the HVDC-linked grids. Specifically, a state feedback controller $\mathbf{r}(t) = -\mathbf{K}\cdot\widetilde{\mathbf{X}}(t)$ is implemented based on the data-driven model (24), (25) to minimize the variations in $f_i$, $f_r$, and $V_{dc}$ while restoring them to the nominal values in the steady state. Considering the control efforts of the HVDC system and generators, a cost function for the optimal SFC is formulated as:

$$\arg\min_{\mathbf{r}(t)=-\mathbf{K}\cdot\widetilde{\mathbf{X}}(t)} J = \int_0^\infty \left( \mathbf{Y}(t)^T \cdot \mathbf{Q} \cdot \mathbf{Y}(t) + \mathbf{r}(t)^T \cdot \mathbf{R} \cdot \mathbf{r}(t) \right) dt, \quad (26)$$

where $\mathbf{Q}$ is a diagonal matrix with weighting coefficients $q_i$ for $1 \leq i \leq z$, each of which is multiplied by the square of each output variable $y_i(t)$ in $\mathbf{Y}(t)$. In other words, the first term in the integral can be expressed equivalently as $\Sigma_i q_i \cdot y_i(t)^2$. Similarly, $\mathbf{R}$ is a weighting-coefficient matrix for $\mathbf{r}(t)$. Note that (26) is formulated using the weighting coefficients for the outputs, unlike in previous studies on optimal system control (e.g., [28]) using weighting coefficients for states. This is because $\widetilde{\mathbf{X}}(t)$ in (24) does not contain information on the physical states of the HVDC-linked grids. For $\mathbf{Q}$ and $\mathbf{R}$, the optimal $\mathbf{K} = \mathbf{R}^{-1}\cdot\widetilde{\mathbf{B}}_\mathbf{r}^T\cdot\mathbf{P}$ exists, such that $\mathbf{P}$ is the solution to a data-driven Riccati equation:

$$\widetilde{\mathbf{A}}^T\cdot\mathbf{P} + \mathbf{P}\cdot\widetilde{\mathbf{A}} + \mathbf{Q} - \mathbf{P}\cdot\widetilde{\mathbf{B}}\cdot\mathbf{R}^{-1}\cdot\widetilde{\mathbf{B}}^T\cdot\mathbf{P} = \mathbf{O}. \quad (27)$$

Note that as shown in (27), the data-driven modeling of the HVDC-linked grids via the OKID and ERA enables the purely data-driven design of the LQ regulator $\mathbf{K}$ using $\widetilde{\mathbf{A}}$ and $\widetilde{\mathbf{B}}$ in (24).

To establish the optimal $\mathbf{r}(t) = -\mathbf{K}\cdot\widetilde{\mathbf{X}}(t)$, all states in $\widetilde{\mathbf{X}}(t)$ need to be calculated using (24) in real time, which increases the computational burden for the optimal SFC in practice. To deal with this issue, a Kalman filter is applied to estimate the unknown, unmeasurable states using the input and output measurement data, as shown in Fig. 6. The data-driven modeling of the HVDC-linked grids also enables data-driven implementation of the Kalman filter using $\widetilde{\mathbf{A}}$, $\widetilde{\mathbf{B}}$, and $\widetilde{\mathbf{C}}$ in (24) and (25). The optimal $\mathbf{r}(t)$ is then established using the data-driven Kalman filter as:

$$\mathbf{r}(t) = -\mathbf{K}\cdot\widetilde{\mathbf{X}}(t) \approx -\mathbf{K}\cdot\widehat{\mathbf{X}}(t), \quad (28)$$

where $\widehat{\mathbf{X}}$ is the estimate of $\widetilde{\mathbf{X}}$.

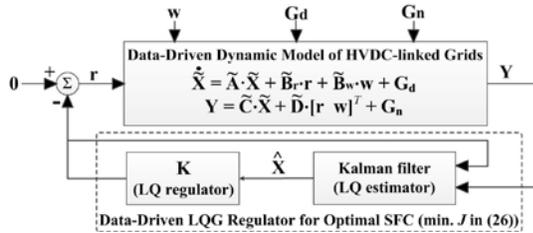

Fig. 6. A data-driven LQG regulator, combining an LQ regulator with a Kalman filter, for optimal SFC in both-side grids linked via the LCC HVDC system.

## IV. CASE STUDIES AND SIMULATION RESULTS

### A. Test System and Simulation Conditions

The proposed strategy for optimal data-driven FR was tested for the parameter sets and operating conditions of two different LCC HVDC systems: i.e., the Jeju-Haenam (JH) system [29] and the CIGRE benchmark system [30]. The JH HVDC system has a rated dc power and voltage of 150 MW and 184 kV, respectively, operating with constant and time-varying power references under normal conditions. The JH system transfers surplus power in the Haenam grid at the rectifier terminal to the Jeju grid, a weak grid with a short circuit ratio of 4.0, at the inverter terminal via a 100-km dc cable. For FR in the Jeju grid, $V_{dcr}$ and $I_{dci}$ are regulated in real time at the rectifier and inverter terminals. Note that in the CIGRE benchmark system, $V_{dci}$ and $I_{dcr}$ are controlled at the inverter and rectifier terminals, respectively. The proposed FR strategy can be well applied to both HVDC systems with little modification.

Table I provides the real parameters of the JH HVDC system, as well as the model parameters of 12 and 8 synchronous generators in the rectifier- and inverter-side grids, respectively. For simplicity, the generators were assumed to operate with the same droop constants and participation factors: i.e., $1/R_{gni(r)} = 1/N_{i(r)}\cdot 1/R_{gi(r)}$ and $\beta_n = 1/N_{i(r)}$ for $1 \leq n \leq N_{i(r)}$, where $N_{i(r)}$ is the number of generators in the inverter- and rectifier-side grids, respectively. Table I also lists the moments of mechanical and emulated inertia and the damping constant for each grid, as well as the PI control gains for the conventional FR strategy.

Using the parameters in Table I, the physics-based dynamic model of the HVDC-linked grids was implemented using both the state-space form (see Appendix) and the MATLAB/SIMULINK (or simply SIMULINK) software, to validate the data-driven, reduced-order model and hence the proposed FR strategy. In particular, the SIMULINK model was constructed using a converter transformer, 12 thyristor valves, and internal firing/extinction angle controllers for each converter. The SIMULINK model also included a library model of a three-phase round-rotor synchronous machine [31], reflecting complicated interactions between the stator and rotor flux linkages.

Table II lists the main features of the proposed (Case 1) and conventional (Cases 2 and 3) strategies for FR support of the LCC HVDC system. In the case studies, the Cases 1 and 2 were compared to validate the effect of the optimal SFC on FR in both-side grids, given the IRE and PFC schemes. Moreover, Case 3 represents a common control strategy for the HVDC system [6], [8] to support only inverter-side FR by regulating $I_{dci}$ and maintaining $V_{dcr}$ as constant: i.e., no feedback loops for the PFC with $1/R_r$, $s \cdot W_r$, $K_r$ and $K_i$ and the SFC with $\Delta V^S_{dcr\_ref}$. A comparison was made between Cases 1 and 3 to confirm the effectiveness of the time-varying control of $V_{dcr\_ref}$ in improving FR in both-sides grids. Fig. 7 shows the profiles of the load demand variations $\Delta P_{li}$ and $\Delta P_{lr}$ in the inverter- and rectifier-side grids, respectively, where the scaled-up RegD signals [32] were reflected over a time period of 200 s. It also shows the variation in the WF power generation $\Delta P_w$ [33] on the rectifier side. In addition to these profiles, stepwise variations in $\Delta P_{li}$ and $\Delta P_{lr}$ were considered in the comparative case studies on Cases 1–3.

TABLE I. PARAMETERS FOR THE CASE STUDIES

| Devices | Parameters | Values | Descriptions |
|---|---|---|---|
| HVDC system | $V_{dcr0}$, $V_{dci0}$ [kV] | 184.0, 183.5 | Nominal dc voltages |
| | $I_{dcr0}$, $I_{dci0}$ [A] | 407.6 | Nominal dc currents |
| | $R_{dc}$ [Ω], $L_{dc}$ [H], $C_{dc}$ [μF] | 1.116, 0.2, 54 | HVDC-link parameters |
| | $X_{cr}$, $X_{ci}$ [Ω] | 7.99 | Converter reactance |
| | $α_0$, $γ_0$ [°] | 15, 18 | Firing, extinction angles |
| | $μ_{i(r)0}$ [°] | 2.44 | Converter overlap angles |
| | $TR_{i(r)}$ | 0.9 | TR tap ratios |
| | kpr, kir | 5.5, 20.1 | Voltage controller gains |
| | kpi, kii | 0.001, 10.0 | Current controller gains |
| | $B_{i(r)}$ | 2 | Number of bridges |
| Syn. machine | $x_d$, $x_d'$, $x_d''$ | 0.2, 0.033, 0.0264 | Coefficients of $dq$-axis flux linkages |
| | $x_q$, $x_q'$, $x_q''$ | 0.19, 0.061, 0.03 | |
| | $T_d'$, $T_d''$, $T_q'$, $T_q''$ | 5.0, 0.05, 0.4, 0.04 | $dq$-axis time constants of rotor flux linkage |
| | $M_s$, $k_d$ | 0.4, 0.001 | Inertia and damping |
| | $ω_{r0}$ | 1.0 | Initial speed of rotor |
| | $N_r$, $N_i$ | 12, 8 | Number of generators |
| | $P_{g0}$, $Q_{g0}$, $V_{g0}$, $θ_{g0}$, [pu] | 1.0, 1.59, 1.01, -1.50 1.0, 1.34, 1.03, 8.21 1.0, 0.00, 1.03, 0.00 1.0, 1.80, 1.01, -10.2 | Initial active and reactive power outputs and terminal bus voltages |
| Governor and gas turbine | $e_g$, $u_g$ | 1.0, 0.05 | Valve positioner gains |
| | $X_g$, $Y_g$ | 0.6, 1.0 | Speed governor gains |
| | $T_{cr}$, $T_f$, $T_{cd}$ | 0.01, 0.23, 0.2 | Time constants of combustion reaction, fuel, compressor discharge |
| Grids | $M_{i(r)}$, $D_{i(r)}$ | 5, 1 | Inertia and damping |
| PFC | $W_i$, $W_r$ | 5 | Emulated inertia |
| | $K_i$, $K_r$, $R_i$, $R_r$ | 0.5 | Droop gains (HVDC) |
| | $R_{gi}$, $R_{gr}$ | 0.5 | Droop gains (Gen.) |
| SFC (Conv.) | $KP_r$, $KP_i$, $KI_r$, $KI_i$ | 3, 3, 25, 25 | PI gains (HVDC) |
| | $KP_r$, $KP_i$, $KI_r$, $KI_i$ | 0.8, 0.8, 0.2, 0.2 | PI gains (Gen.) |

TABLE II. FEATURES OF THE PROPOSED AND CONVENTIONAL FR STRATEGIES

| FR strategies | | Model | $\Delta V_{dc\_ref}$ | FR | SFC |
|---|---|---|---|---|---|
| Proposed | Case 1 | data-driven | time-varying | both grids | LQG |
| Conventional | Case 2 | data-driven | time-varying | both grids | PI |
| | Case 3 | physics-based | fixed | inverter grid | PI |

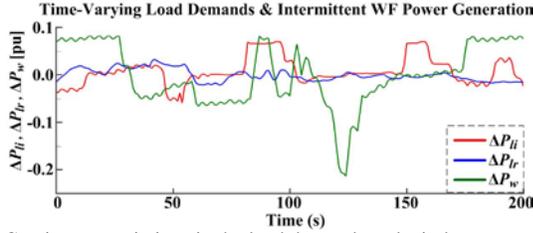

Fig. 7. Continuous variations in the load demands and wind power generation.

### B. Verification of Data-driven Model of HVDC-linked Grids

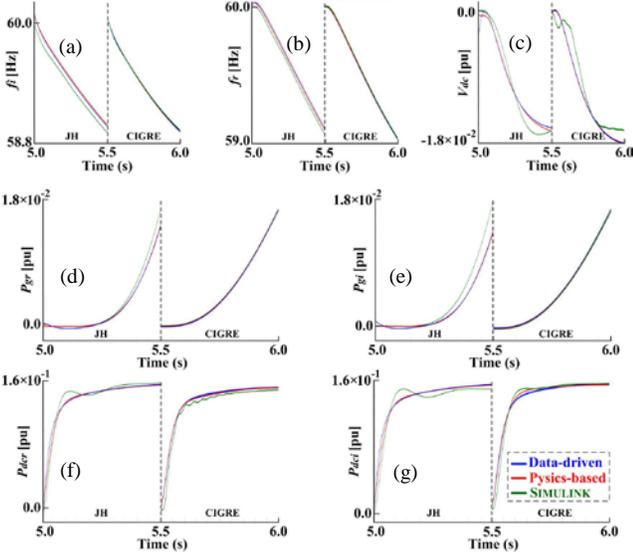

Fig. 8. Comparisons between the step responses of the data-driven, reduced-order model, the physics-based full-order model, and the comprehensive SIMULINK model of the HVDC-linked grids to $\Delta P_{li}(t) = 0.3$ pu: (a) $f_i$, (b) $f_r$, (c) $V_{dc}$, (d) $P_{gr}$, (e) $P_{gi}$, (f) $P_{dcr}$, and (g) $P_{dci}$.

Fig. 8 shows comparisons of the dynamic responses of the three different models of the HVDC-linked grids to a step increase in $P_{li}$ by 0.3 pu: i.e., the data-driven, reduced-order model, the physics-based, full-order model, and the comprehensive SIMULINK model, discussed in Sections III and IV-A, and in the Appendix. Fig. 8(a)–(c) show that, for both the JH and CIGRE HVDC systems, the dynamic responses of the three models were very similar for the profiles of $f_i$, $f_r$, and $V_{dc}$. Fig. 8(d)–(g) also show good consistency between the dynamic models for the total generated power profiles $P_{gr}$ and $P_{gi}$ and the dc transferred power profiles $P_{dcr}$ and $P_{dci}$. This verifies that the data-driven model successfully reflected the dynamic operating characteristics of the HVDC-linked grids, thus demonstrating the enhanced practical applicability of the proposed FR strategy. The comparisons also validated the accuracy of the case study results acquired using the data-driven model, as discussed in Sections IV-C, D, and E.

### C. Proposed FR Strategy for Stepwise Load Variations

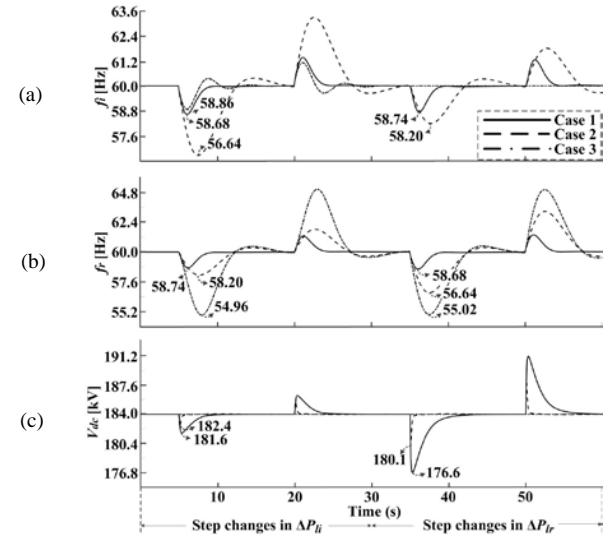

Fig. 9. Step responses to $\Delta P_{li}$ ($t = 5^+$ s) = $\Delta P_{lr}$ ($t = 35^+$ s) = 0.3 pu during 15 s for the proposed and conventional FR strategies: (a) $f_i$, (b) $f_r$, and (c) $V_{dc}$.

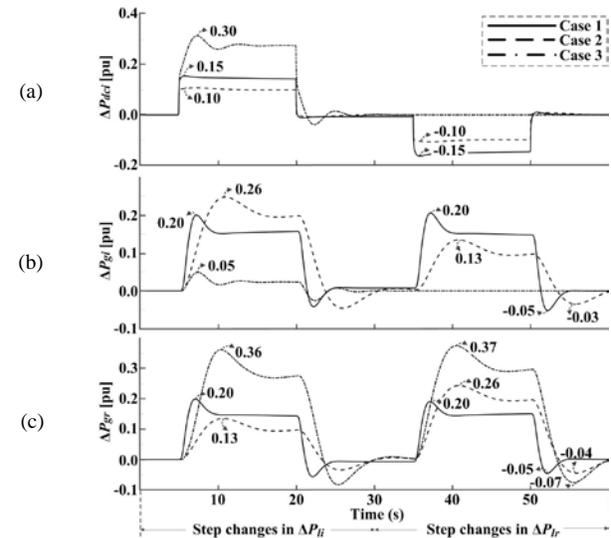

Fig. 10. Corresponding variations in the dc transferred and ac generated power: (a) $\Delta P_{dci}$, (b) $\Delta P_{gi}$, and (c) $\Delta P_{gr}$.

TABLE III. COMPARISONS OF THE STEP RESPONSE TEST RESULTS

| Max. [Hz], [pu] | (1) $\|\Delta f_i\|$ | (2) $\|\Delta f_r\|$ | (1)+(2) | (3) $\|\Delta P_{dci}\|$ | (4) $\|\Delta P_{dcr}\|$ | (3)+(4) | (5) $\|\Delta P_{gi}\|$ | (6) $\|\Delta P_{gr}\|$ | (5)+(6) |
|---|---|---|---|---|---|---|---|---|---|
| Case 1 | 1.32 | 1.32 | 2.64 | 0.15 | 0.15 | 0.30 | 0.20 | 0.20 | 0.40 |
| Case 2 | 3.36 | 3.36 | 6.72 | 0.10 | 0.10 | 0.20 | 0.26 | 0.26 | 0.52 |
| Case 3 | 1.14 | 5.04 | 6.18 | 0.30 | 0.30 | 0.60 | 0.05 | 0.37 | 0.42 |

Fig. 9 shows the dynamic responses of the both-side grids connected via the JH HVDC system to step increases in $\Delta P_{li}$ and $\Delta P_{lr}$ of 0.3 pu at $t = 5$ s and 35 s, respectively, during a period of 15 s. As specified in Table III, the proposed FR strategy (i.e., Case 1) decreased the sum of the maximum deviations of $f_i$ and $f_r$ (i.e., $|\Delta f_i|_{max}+|\Delta f_r|_{max}$) by 60.7% and 57.3%, compared to the conventional strategies (i.e., Cases 2 and 3, respectively), while also resulting in a maximum variation in $V_{dc}$ of 4.0%. Moreover, in Case 1, $f_i$ and $f_r$ were restored back to the nominal value more rapidly, and with a smaller overshoot than in Cases 2 and 3. This confirmed that the data-driven LQG regulator for the optimal SFC, discussed in Section III-C, successfully improved the frequency stability compared to the PI controllers for the conventional SFC.

Fig. 10 represents the corresponding profiles of $\Delta P_{dci}$, $\Delta P_{gi}$, and $\Delta P_{gr}$. In Case 1, both $\Delta P_{gi}$ and $\Delta P_{gr}$ were increased by up to 0.20 pu and then maintained at approximately 0.15 pu after $t = 10$ s, when $\Delta P_{li}$ was increased by 0.30 pu. The HVDC system immediately increased $\Delta P_{dci(r)}$ by 0.15 pu and maintained it at an almost constant level to transfer the increased power generation in the rectifier-side grid to the inverter-side grid, to effectively support real-time FR in both the transient and steady states. In other words, the proposed coordinated control of the HVDC system and generators via the data-driven LQG regulator enabled the load demand variation to be shared optimally between the generators in both-side grids, reducing the deviations and overshooting of $f_i$ and $f_r$. The HVDC system and generators operated in a similar way under an increase in $\Delta P_{lr}$ of 0.30 pu at $t = 35$ s. In Case 2, the PI controllers led the generators to compensate for the load variations primarily on the same side of the HVDC system. For $\Delta P_{li}(t = 5^+$ s$) = 0.30$ pu, $\Delta P_{gi}$ increased by up to 0.26 pu, whereas $\Delta P_{gr}$ increased by 0.13 pu. The HVDC system transferred only $\Delta P_{dci(r)} = 0.10$ pu to the inverter-side grid, resulting in larger deviations in $f_i$ and $f_r$ in the transient state. In Case 3, $\Delta P_{li}(t = 5^+$ s$) = 0.3$ pu caused larger variations in $P_{gr}$ and $f_r$ than in $P_{gi}$ and $f_i$, because only inverter-side FR was taken into consideration. It can be problematic when the rectifier-side grid contains critical loads and generators with limited reserve capacities. In all cases, $\Delta P_{dci}$ and $\Delta P_{dcr}$ changed faster than $\Delta P_{gi}$ and $\Delta P_{gr}$, verifying the effectiveness of the IRE and PFC schemes in exploiting the fast response of the HVDC converters.

### D. Proposed FR Strategy for Continuous Load Variations

The proposed FR strategy was also tested for continuous variations in $\Delta P_{li}$, $\Delta P_{lr}$, and $\Delta P_w$, as shown in Fig. 7. Fig. 11 shows the corresponding profiles of $f_i$, $f_r$, and $V_{dc}$, and Table IV lists the rms variations of $\Delta f$ and $\Delta P_g$ in the inverter- and rectifier-side grids, estimated as:

$$\Delta f_{rms} = \left(\sum_{s=1}^{S} \Delta f_s^2 / S\right)^{1/2} \quad \text{and} \quad \Delta P_{g,rms} = \left(\sum_{s=1}^{S}\left(\sum_{n=1}^{N_{i(r)}} \Delta P_{gni(r),s}^2\right)/S\right)^{1/2}. \quad (29)$$

In (29), $S$ is the total number of measurement samples over a period of 200 s and $s$ is the index of the sample. In Case 1, the sum of $\Delta f_{i,rms}$ and $\Delta f_{r,rms}$ was reduced by 73.5% and 71.9%, compared to those in Cases 2 and 3, respectively, validating the effectiveness of the proposed HVDC system control for supporting FR in both-side grids. Furthermore, $\Delta P_{g,rms}$ in Case 1 was 22.2% and 30.0% smaller than in Cases 2 and 3, respectively. This is mainly because the real-time control of the HVDC system and the synchronous generators on both sides were optimally coordinated to compensate for the total variation in the load demand and WF power generation more effectively. The case study results also imply that the proposed control can reduce the operating requirements (e.g., reserve capacity) of the generators. The cost due to the increased operational stress of the HVDC system can be compensated for by the savings in operating costs of the generators resulting from the improved stability and flexibility of real-time FR.

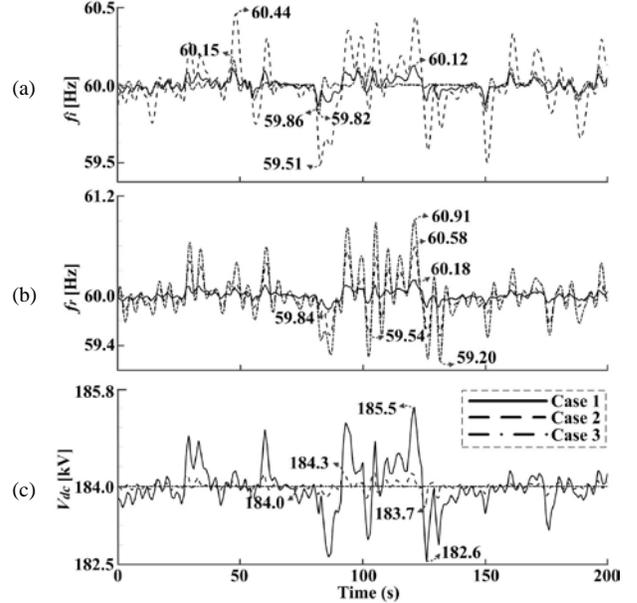

Fig. 11. Dynamic responses of the HVDC-linked grids to continuous variations of $\Delta P_{li}$, $\Delta P_{lr}$, and $\Delta P_w$ for the proposed and conventional FR strategies: (a) $f_i$, (b) $f_r$, and (c) $V_{dc}$.

TABLE IV. COMPARISONS OF THE CONTINUOUS RESPONSE TEST RESULTS

| $\Delta f_{rms}$ and $\Delta P_{g,rms}$ [Hz] [pu] | (1) $\Delta f_{i,rms}$ | (2) $\Delta f_{r,rms}$ | (1)+(2) | (3) $\Delta P_{gi,rms}$ | (4) $\Delta P_{gr,rms}$ | (3)+(4) |
|---|---|---|---|---|---|---|
| Case 1 | 0.04 | 0.05 | 0.09 | 0.03 | 0.04 | 0.07 |
| Case 2 | 0.17 | 0.17 | 0.34 | 0.04 | 0.05 | 0.09 |
| Case 3 | 0.04 | 0.28 | 0.32 | 0.01 | 0.09 | 0.10 |

### E. Performance Evaluations under Various Conditions

The case studies discussed in Sections IV-C and IV-D were repeated under various operating conditions of the HVDC system. In the proposed FR strategy, the HVDC system successfully supported real-time FR in both-side grids for different values of the LQG parameters $\mathbf{Q}$ and $\mathbf{R}$ in (26). For example, Table V shows the results of the step response test of the HVDC-linked grids, when the weighting coefficients in $\mathbf{Q}$ were increased by 10 times compared to those used in Section IV-C. From Fig. 9 and Tables III and V, it can be seen that the maximum variations in $f_i$, $f_r$, and $V_{dc}$ for Case 1 became smaller than those for Case 2. In other words, the proposed strategy improved the transient stability of both the HVDC system and the interfacing grids. Note that, in practice, $\mathbf{Q}$ should be determined considering the load composition and installed capacity of the synchronous generators and wind turbines. Moreover, Table VI shows that the values of $\Delta f_{i,rms}+\Delta f_{r,rms}$ and $\Delta P_{gi,rms}+\Delta P_{gr,rms}$ in Case 1 were smaller than in Cases 2 and 3, under all different conditions of the IRE and PFC schemes of the HVDC system. The case studies were also performed using the CIGRE benchmark system with a rated power and voltage of 1 GW and 500 kV, respectively. The proposed strategy was still effective in improving the frequency stability and operating efficiency in both-side grids under different HVDC specifications, as shown in Table VII.

TABLE V. STEP RESPONSE TEST RESULTS FOR DIFFERENT VALUES OF **Q** IN (26)

| $|\Delta f_{i(r)}|$ [Hz] | (1) $|\Delta f_i|_{max}$ | (2) $|\Delta f_r|_{max}$ | (1)+(2) | $|\Delta V_{dc}|_{max}$ [kV] | for $\Delta P_{li}$ | for $\Delta P_{lr}$ |
|---|---|---|---|---|---|---|
| | 1.20 | 1.20 | 2.40 | | 0.55 | 1.84 |
| $|\Delta P_{dci}|$ [pu] | (3) $|\Delta P_{dci}|_{max}$ | (4) $|\Delta P_{dcr}|_{max}$ | (3)+(4) | $|\Delta P_{gi}|$ [pu] | (5) $|\Delta P_{gi}|_{max}$ | (6) $|\Delta P_{gr}|_{max}$ | (5)+(6) |
| | 0.16 | 0.16 | 0.32 | | 0.17 | 0.17 | 0.34 |

TABLE VI. CONTINUOUS RESPONSE FOR DIFFERENT IRE AND PFC CONDITIONS

| $\Delta f_{rms}$ and $\Delta P_{g,rms}$ [Hz] [pu] | | (1) $\Delta f_{i,rms}$ | (2) $\Delta f_{r,rms}$ | (1)+(2) | (3) $\Delta P_{gi,rms}$ | (4) $\Delta P_{gr,rms}$ | (3)+(4) |
|---|---|---|---|---|---|---|---|
| a | Case 1 | 0.04 | 0.06 | 0.10 | 0.04 | 0.04 | 0.08 |
| | Case 2 | 0.18 | 0.18 | 0.36 | 0.05 | 0.05 | 0.10 |
| | Case 3 | 0.06 | 0.27 | 0.33 | 0.02 | 0.09 | 0.11 |
| b | Case 1 | 0.05 | 0.05 | 0.11 | 0.04 | 0.05 | 0.09 |
| | Case 2 | 0.20 | 0.26 | 0.46 | 0.05 | 0.06 | 0.11 |
| | Case 3 | 0.12 | 0.28 | 0.40 | 0.03 | 0.09 | 0.12 |

a: no PFC, b: neither IRE nor PFC

TABLE VII. CONTINUOUS RESPONSE FOR THE CIGRE BENCHMARK SYSTEM

| $\Delta f_{rms}$ and $\Delta P_{g,rms}$ [Hz] [pu] | (1) $\Delta f_{i,rms}$ | (2) $\Delta f_{r,rms}$ | (1)+(2) | (3) $\Delta P_{gi,rms}$ | (4) $\Delta P_{gr,rms}$ | (3)+(4) |
|---|---|---|---|---|---|---|
| Case 1 | 0.04 | 0.05 | 0.09 | 0.02 | 0.05 | 0.07 |
| Case 2 | 0.17 | 0.15 | 0.32 | 0.03 | 0.06 | 0.09 |
| Case 3 | 0.04 | 0.26 | 0.30 | 0.01 | 0.08 | 0.09 |

## V. CONCLUSIONS

This paper proposed an optimal data-driven control strategy for an LCC HVDC system, in which the dc voltage and current were optimally regulated at distinct HVDC terminals to improve real-time FR in both the rectifier- and inverter-side grids. In the proposed strategy, each HVDC converter was implemented with feedback loops for droop control, as well as inertia emulation, thereby stabilizing the grid frequency and dc voltage in a localized manner. For the optimal FR, a data-driven model of the HVDC-linked grids was then developed based on input-output data measurements to design a data-driven LQG controller, which was integrated with the converter feedback loops. This enabled the HVDC system to support frequency stability via the inertia emulation, PFC, and SFC in coordination with the synchronous generators. Simulation case studies were conducted using the data-driven models on the real HVDC system and CIGRE benchmark system, which were verified via comparisons with the physics-based and comprehensive SIMULINK models. The proposed optimal data-driven control strategy decreased the maximum frequency variation by 60.7% and 57.3% for the step load variation and the rms variation by 73.5% and 71.9% for the continuous load variation, compared to the conventional FR strategies. The optimal coordination of the HVDC system and generators effectively reduced the variations in frequency and thermal power generation under various HVDC system specifications, LQG parameters, and inertia emulation and droop control conditions.

## REFERENCES

[1] D. Xiang, L. Ran, and J. R. Bumby *et al.*, "Coordinated control of an HVDC link and doubly fed induction generators in a large offshore wind farm," *IEEE Tans. Power Del.*, vol. 21, no. 1, pp. 463–471, Jan. 2006.
[2] S. Favuzza *et al.*, "New energy corridors in the Euro-Mediterranean area: the pivotal role of Sicily," *Energies*, vol. 11, no. 6, pp. 1–14, 2018.
[3] N. Anderson, "Shetland HVDC link project", Scottish and Southern Energy, Aug. 2016. [Online]. Available: https://www.ssen-transmission.co.uk/media/1514/shetland-hvdc-link-consultation-summary-booklet-august-2016.pdf
[4] Midwest ISO, "MISO Transmission Planning Processes," Tech. Rep. 2017. [Online]. Available: https://www.cne.cl/wp-content/uploads/2017/04/Dale-Osborn.pdf
[5] R. Li *et al.*, "Frequency control design for offshore wind farm grid with LCC-HVDC link connection," *IEEE Trans. Power Electron.*, vol. 23, no. 3, pp. 1085–1092, 2008.
[6] Z. Miao *et al.*, "Wind farms with HVdc delivery in inertial response and primary frequency control," *IEEE Trans. Energy Convers.*, vol. 25, no. 4, pp. 1171–1178, 2010.
[7] A. Yogarathinam *et al.*, "Impact of inertia and effective short circuit ratio on control of frequency in weak grids interfacing LCC-HVDC and DFIG-based wind farms," *IEEE Trans. Power Del.*, vol. 32, no. 4, pp. 2040–2051, 2017.
[8] L. Wang and M. Sa-Nguyen Thi, "Stability enhancement of a PMSG-based offshore wind farm fed to a multi-machine system through an LCC-HVDC link," *IEEE Trans. Power Syst.*, vol. 28, no. 3, pp. 3327–3334, Aug. 2013.
[9] S. G. Vennelaganti *et al.*, "New insights into coupled frequency dynamics of AC grids in rectifier and inverter sides of LCC-HVDC interfacing DFIG-based wind farms," *IEEE Tran. Power Del.*, vol. 33, no. 4, pp. 1765–1776, Aug. 2018.
[10] S. P. Azad, R. Iravani, and J. E. Tate, "Damping inter-area oscillations based on a model predictive control (MPC) HVDC supplementary controller," *IEEE Trans. Power Syst.*, vol. 28, no. 3, pp. 3174–3183, Aug. 2013.
[11] S. P. Azad *et al.*, "Decentralized supplementary control of multiple LCC-HVDC links," *IEEE Trans. Power Syst.*, vol. 31, no. 1, pp. 572–580, Jan. 2016.
[12] A. Gustafsson, M. Saltzer *et al.*, "The new 525 kV extruded HVDC cable system," ABB Grid Systems, Tech, Rep. Aug. 2014.
[13] *IEEE Guide for Test Procedures for HVDC Thyristor Valves*, IEEE Standard 857-1990, pp. 1-24, Feb. 1990.
[14] D. Kwon *et al.*, "Modeling and analysis of an LCC HVDC system using DC voltage control to improve transient response and short-term power transfer capability," *IEEE Trans. Power Del.*, vol. 33, no. 4, pp. 1922–1933, Aug. 2018.
[15] Y. Xue *et al.*, "Reactive power and AC voltage control of LCC HVDC system with controllable capacitors," *IEEE Trans. Power Syst.*, vol. 32, no. 1, pp. 753–764, 2017.
[16] "Sensing and measurement," National Energy Technology Lab., 2007. [Online]. Available: http://www.smartgrid.gov/files/appendix_b2_sensing_measurement.pdf
[17] S. Martin-Martinez *et al.*, "Wind power variability and singular events," in *Advances in Wind Power*, Janeza, Croatia, 2012, ch 12, pp. 285–304.
[18] H. Zhang, J. Zhou, and Q. Sun *et al.*, "Data-driven control for interlinked AC/DC microgrids via model-free adaptive control and dual-droop control," *IEEE Trans. Smart Grid*, vol. 8, no. 2, pp. 557–571, Dec. 2015.
[19] Y. Xia *et al.*, "Adaptive-observer-based data driven voltage control in islanded-mode of distributed energy resource systems," *Energies*, vol. 11, pp. 1–14, Nov. 2018.
[20] Z. Yan and Y. Xu, "Data-driven load frequency control for stochastic power systems: a deep reinforcement learning method with continuous action search," *IEEE Trans. Power Syst.*, vol. 34, no. 2, pp. 1653–1656, Nov. 2018.
[21] L. Yin, T. Yu, and L. Zhou *et al.*, "Artificial emotional reinforcement learning for automatic generation control of large-scale interconnected power grids," *IET Gener., Transm., Distrib.*, vol. 11, no. 9, pp. 2305–2313, Jun. 2017.
[22] J. N. Juang, M. Phan, and L. G. Horta *et al.*, "Identification of observer/Kalman filter Markov parameters: Theory and experiments," in Proc. *AIAA Guidance, Navigation, Contr. Conf.*, New Orleans, LA, 1991, pp. 1172–1179.
[23] J. N. Juang and R. S. Pappa, "An eigensystem realization algorithm for modal parameter identification and model reduction," *J. Guidance, Contr. Dynamics*, vol. 8, no. 5, pp. 620–627, Oct. 1985.
[24] K. P. S. Parmar, "State space based load frequency control of multi-area power systems," Ph.D. dissertation, Dept. Electon., Eng., Indian Institute of Technology Guwahati, 2013.
[25] R. E. Kalman, "A new approach to linear filtering and prediction problems," *J.Basic Eng.*, vol. 82, no. 1, pp. 35–45, Mar. 1960.
[26] S. L. Brunton and J. N. Kutz, *Data-driven science and Engineering*, Cambridge: Cambridge University Press, 2018, pp. 321–343.
[27] L. S. Shieh, H. Wang, and R. E. Yates, "Discrete-continuous model conversion," *Appl. Math. Model*, vol. 4, no. 6, pp. 449–455, Dec. 1980.
[28] A. Al-Digs *et al.*, "Measurement-based sparsity-promoting optimal control of line flows," *IEEE Trans. Power Syst.,* vol. 33, no. 5, pp. 5628–5638, Sep. 2018.
[29] D. Yoon *et al.*, "Smart operation of HVDC systems for large penetration of wind energy resources," *IEEE Trans. Smart Grid*, vol. 4, no. 1, pp. 359–366, Mar. 2013.
[30] H. Atighechi *et al.*, "Dynamic Average-Value Modeling of CIGRE HVDC Benchmark System," *IEEE Trans. Power Del.*, vol. 29, no. 5, pp. 2046–2054, 2014.
[31] Mathworks, *Synchronous Machine*. [Online]. Available: https://kr.mathworks.com/help/physmod/sps/powersys/ref/synchronousmachine.html
[32] Fast Response Regulation Signal, PJM [Online]. Available: http://www.pjm.com/markets-and-operations/ancillary-services/mkt-based-regulation/fast-response-regulation-signal.aspx
[33] Y. Wan, "Wind Power Plant Behaviors: Analysis of Long-Term Wind Power Data," National Renewable Energy Lab., Golden, CO, Tech. Rep.NREL/TP-500-36651, 2004. [Online]. Available: http://www.nrel.gov/docs/fy04osti/36551.pdf
[34] K. N. Shubhanga and Y. Ananthollа, Manual for a multi-machine small signal stability programme. Surathkal: NITK 2009. [Online] Available:http://www.ee.iitb.ac.in/~anil/download.